\documentclass[lettersize,journal]{IEEEtran}
\usepackage{amsmath,amsfonts}
\usepackage{algorithmic}
\usepackage{algorithm}
\usepackage{array}
\usepackage[caption=false,font=normalsize,labelfont=sf,textfont=sf]{subfig}
\usepackage{textcomp}
\usepackage{stfloats}
\usepackage{url}
\usepackage{verbatim}
\usepackage{graphicx}
\usepackage{cite}
\usepackage{multirow}
\usepackage{cleveref}
\usepackage{xspace}
\usepackage{bbm}
\usepackage{booktabs}
\usepackage{color}

\usepackage[ruled,linesnumbered,algo2e]{algorithm2e}
\usepackage[framemethod=TikZ]{mdframed}
\usepackage{amsthm}
\newcounter{prf}[section]
\renewcommand{\theprf}{\arabic{section}.\arabic{prf}}

\newcommand{\ourmethod}{\texttt{GS-NP}\xspace}
\begin{document}

\title{Graph Stochastic Neural Process for Inductive Few-shot Knowledge Graph Completion}

\author{Zicheng Zhao, Linhao Luo, Shirui Pan, \textit{Senior Member, IEEE}, \\ Chengqi Zhang, \textit{Senior Member, IEEE}, and Chen Gong, \textit{Senior Member, IEEE}

\thanks{Z. Zhao and C. Gong are with the Key Laboratory of Intelligent Perception and Systems for High-Dimensional Information, Ministry of Education, School of Computer Science and Engineering, Nanjing University of Science and Technology, Nanjing 210094, China. Email: \{zicheng.zhao, chen.gong\}@njust.edu.cn.}
\thanks{L. Luo is with the Department of Data Science and AI, Monash University, Melbourne, Australia. E-mail: linhao.luo@monash.edu.}
\thanks{S. Pan is with the School of Information and Communication Technology and Institute for Integrated and Intelligent Systems (IIIS), Griffith University, Queensland, Australia. Email: s.pan@griffith.edu.au.}
\thanks{C. Zhang is with the School of Computer Science and Technology, University of Technology Sydney, Sydney 101408, Australia. E-mail: Chengqi.Zhang@uts.edu.au.}
\thanks{Z. Zhao and L. Luo contributed equally to this work.}
\thanks{Corresponding Author: C. Gong.}
}

\markboth{Journal of \LaTeX\ Class Files,~Vol.~14, No.~8, August~2021}%
{Shell \MakeLowercase{\textit{et al.}}: A Sample Article Using IEEEtran.cls for IEEE Journals}

\IEEEpubid{0000--0000/00\$00.00~\copyright~2021 IEEE}

\maketitle
\begin{abstract}
Knowledge graphs (KGs) store enormous facts as relationships between entities. Due to the long-tailed distribution of relations and the incompleteness of KGs, there is growing interest in few-shot knowledge graph completion (FKGC). Existing FKGC methods often assume the existence of all entities in KGs, which may not be practical since new relations and entities can emerge over time. Therefore, we focus on a more challenging task called inductive few-shot knowledge graph completion (I-FKGC), where both relations and entities during the test phase are unknown before. Inspired by the idea of inductive reasoning, we cast I-FKGC as an inductive reasoning problem. Specifically, we propose a novel Graph Stochastic Neural Process approach (\ourmethod), which consists of two major modules. In the first module, to obtain a generalized hypothesis (\textit{e.g.,} shared subgraph), we present a neural process-based hypothesis extractor that models the joint distribution of hypothesis, from which we can sample a hypothesis for predictions. In the second module, based on the hypothesis, we propose a graph stochastic attention-based predictor to test if the triple in the query set aligns with the extracted hypothesis. Meanwhile, the predictor can generate an explanatory subgraph identified by the hypothesis. Finally, the training of these two modules is seamlessly combined into a unified objective function, of which the effectiveness is verified by theoretical analyses as well as empirical studies. Extensive experiments on three public datasets demonstrate that our method outperforms existing methods and derives new state-of-the-art performance.


\end{abstract}


\section{Introduction}
Knowledge graphs (KGs) \cite{9416312}, which contain factual knowledge in terms of structural triples, \textit{i.e.}, \texttt{(head entity, relation, tail entity)},  are widely utilized in various downstream applications, such as recommender systems~\cite{10530439}, community detection \cite{luo2021detecting}, and web search \cite{luo2023gsim}. However, KGs often suffer from incompleteness, limiting their practical applicability. Many approaches have been proposed to complete missing links by inferring from observed facts \cite{bordes2013translating}, \cite{wan2021reasoning}. However, these methods require sufficient training data for each relation, which is not always available. In other words, in real-world KGs, there is a significant amount of presented relations that are only associated with a couple of facts in KGs. This poses a great challenge for predicting missing facts within the KG. To tackle this issue, several few-shot knowledge graph completion (FKGC) methods \cite{luo2023normalizing}, \cite{chen2019meta}, \cite{niu2021relational}, \cite{xiong2018one}, \cite{10154576} have been proposed to predict missing facts for relations with limited supporting triples. 
\begin{figure}[t]
    \centering
    \includegraphics[width=.9\linewidth]{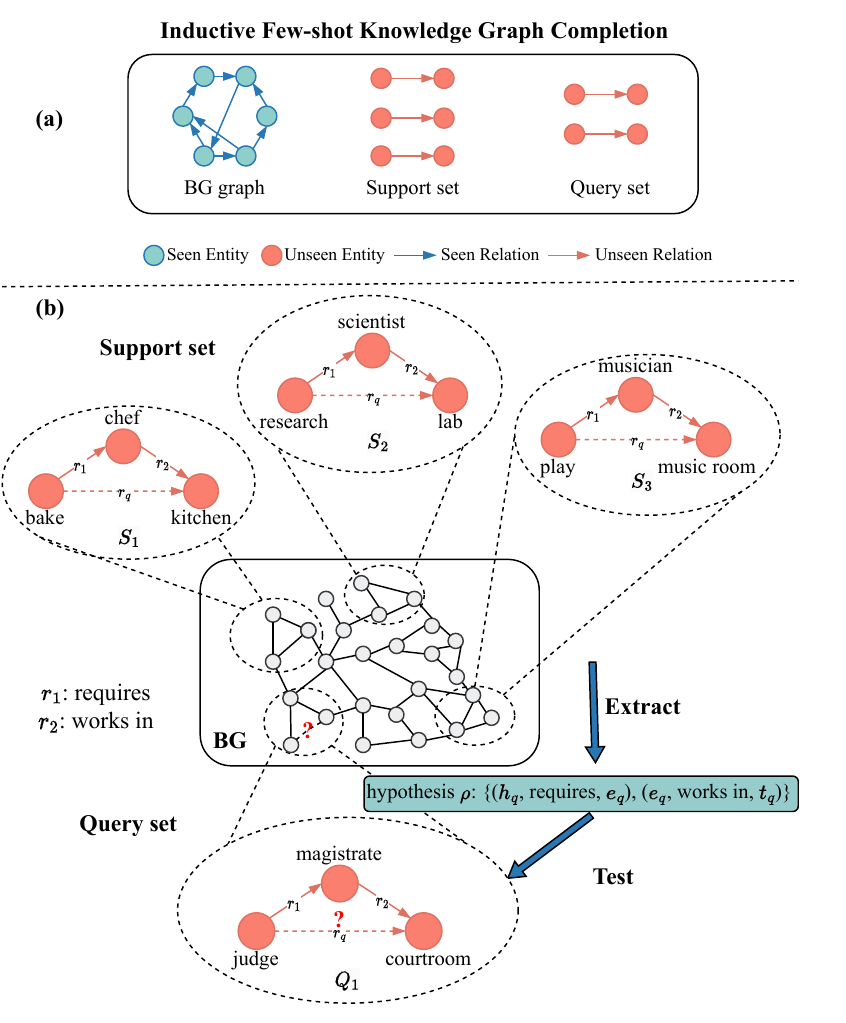}
    \caption{(a) An illustration of inductive few-shot link prediction (I-FKGC) where new relations emerge with unseen entities simultaneously. (b) An example of the motivation behind our method where we try to test whether the query set aligns with the hypothesis extracted from the support set.}
    \label{fig:introduction}
\end{figure}
\IEEEpubidadjcol
Existing few-shot knowledge graph completion (FKGC) methods typically follow the meta-learning framework \cite{hospedales2021meta}, \cite{9791434}. In this framework, the model is either fine-tuned \cite{chen2019meta}, \cite{niu2021relational} on a \textit{support set}, which contains a few triples related to a relation, 
or designs a matching network to calculate the similarity between \textit{query set} (which consists of triples to be predicted) and \textit{support set} for the few-shot task \cite{xiong2018one, zhang2020few, sheng2020adaptive}. Despite the success, these methods often suffer from out-of-distribution (OOD) \cite{huang2022few} and overfitting problems \cite{dong2020mamo}, which cannot extract a general pattern to derive new knowledge effectively. Moreover, they do not consider the dynamic nature of KGs, where new entities emerge with new relations simultaneously. As shown in Figure~\ref{fig:introduction}(a), relations and entities involved in the testing phase are all unseen to the original background knowledge graph (BG Graph). In light of their limitations, we focus on a more challenging setting, \textit{i.e.,} \textbf{inductive few-shot knowledge graph completion (I-FKGC)}, where the model is required to predict missing facts for unseen relations and entities in the query set with limited supporting triples in the support set \cite{huang2022few, liu2023graph}. 

The objective of I-FKGC is to incorporate new relations and new entities into KGs, which requires the model to exhibit strong inductive reasoning ability, enabling it to generalize effectively in unseen scenarios. Inspired by inductive reasoning \cite{DanHunter}, in this paper, we cast I-FKGC as an inductive reasoning problem. We attempt to find a hypothesis (\textit{e.g.}, a pattern or a subgraph) shared among all triples in the support set and then test whether each triple in the query set aligns with this hypothesis for prediction.
%
More specifically, we try to extract a shared hypothesis, as depicted in Figure~\ref{fig:introduction}(b), by examining the subgraphs surrounding the triples in the support set. By analyzing these subgraphs, we can observe that they exhibit similar structures, such as \texttt{(bake, requires, chef), (chef, works in, kitchen)}
and \texttt{(play, requires, musician)}, \texttt{(musician, works in, music room)}. 
Consequently, we can extract a hypothesis: $\rho$: \{\texttt{($h_q$, requires, $e_q$),  ($e_q$, works in, $t_q$)}\}, which explains how to establish the relation $r_q$ between the head entities and tail entities. For instance, bake requires the chef and the chef works in the kitchen. Based on this hypothesis, we can infer that it is highly likely for \texttt{(judge, $r_q$, courtroom)} from the query set to be true since it aligns with our hypothesis, \textit{i.e.,} judge requires magistrate and magistrate works in courtroom.

Based on the above consideration, in this paper, we propose a novel Graph Stochastic Neural Process approach (\ourmethod), which contains two major modules to effectively extract and apply the hypothesis for addressing the I-FKGC task. Existing methods extract the hypothesis by identifying the most commonly shared structure among the support set \cite{huang2022few}. However, due to the limited number of support triples, the extracted hypothesis may not generalize effectively to the query set. Neural process (NP) \cite{garnelo2018neural} offers a new way to deal with limited data by modeling a distribution over the prediction function. Motivated by this, we propose a \emph{neural process-based hypothesis extractor} that treats the hypothesis as a latent variable and models its joint distribution instead of directly extracting it. This approach helps to alleviate overfitting and out-of-distribution issues. By sampling from this distribution, we can obtain the hypothesis for predictions.
Then, we propose a \emph{graph stochastic attention-based predictor} to test whether the query set aligns with the extracted hypothesis. Since the sampled hypothesis is a latent variable, we employ the stochastic attention mechanism \cite{shankar2019posterior} that transforms the hypothesis into a soft edge mask and identifies a subgraph used for predictions. In this way, our method can generalize well to the query set under few-shot scenarios and provide explanations for predictions.
The training of these two modules is seamlessly combined into a unified objective function, of which the effectiveness is verified by both theoretical analyses and empirical studies. 

The main contributions of our paper can be summarized as follows:
\begin{itemize}
    \item We propose a novel neural process approach for inductive few-shot knowledge graph completion. To the best of our knowledge, this is the first work to develop a neural process framework for I-FKGC.
    \item We propose a graph stochastic attention-based predictor to test whether the query set shares the same hypothesis extracted from the support set. Based on the extracted hypothesis, our method can generalize well to the query set under few-shot scenarios and provide explanations for predictions.
    \item We conduct extensive experiments on three typical datasets. Experimental results show that our model (\ourmethod) outperforms existing methods, verifying the effectiveness of our method.
\end{itemize}

The structure of the paper is organized as follows: Section~\ref{sec:related} reviews representative prior works on few-shot knowledge graph completion along with two key techniques integral to our methods. Section~\ref{sec:pre} provides a brief introduction to the concepts of neural processes and graph stochastic attention, and outlines the settings and common notations used throughout the paper. Section~\ref{sec:approach} elaborates on the motivation behind our research and details our proposed framework. Section~\ref{sec:exp} discusses the experimental evaluations and case studies conducted to validate our methods. Section~\ref{sec:con} concludes the paper and discusses potential future research directions.

\section{Related work}
\label{sec:related}
In this section, we review some representative prior works and two key techniques related to this paper.
\subsection{Few-shot Knowledge Graph Completion}
Existing few-shot knowledge graph completion (FKGC) methods can be roughly divided into two groups: meta-optimization-based and metric learning-based models. Meta-optimization-based methods attempt to quickly update the meta parameters via gradient descent on the few-shot data, making models seamlessly generalize to new relations. For instance, MetaR \cite{chen2019meta} proposes a relation-meta learner to extract relation-specific meta-information and then applies it to few-shot relational predictions. GANA \cite{niu2021relational} introduces a gated and attentive neighbor aggregator to improve the quality of generated embeddings and the relation meta representation. However, meta-optimization-based methods are often sensitive to the quality of given few-shot data and suffer from out-of-distribution problems \cite{liu2023good}. Metric learning-based methods try to develop a matching network, which first encodes the triples in the support set and then measures the similarity between the query set and the support set. GMatching \cite{xiong2018one} proposes a neighbor encoder and LSTM matching network to measure the similarity, which is the first work to address the FKGC problem. FSRL \cite{zhang2020few} builds on the neighbor information aggregation mechanism of GMatching by introducing a fixed attention mechanism to consider multiple support triples. FAAN \cite{sheng2020adaptive} extends FSRL by presenting a relation-specific adaptive neighbor encoder.

To handle unseen entities, the problem of inductive link prediction has also received much attention \cite{teru2020inductive}, \cite{wang2021relational}, \cite{xu2022subgraph}, \cite{zhao2023towards}. GraIL \cite{teru2020inductive} utilizes the enclosing subgraph around the target triple for inductive link prediction. PatchCon \cite{wang2021relational} introduces a novel relation message-passing mechanism to capture the inductive features. However, these methods can only generalize to graphs with new entities while cannot handle unseen relations \cite{galkin2023towards}. To address this issue, CSR \cite{huang2022few} combines the merit of existing FKGC and inductive link prediction methods, introducing a novel subgraph-based matching network. It tries to extract connection subgraphs among the support triples and tests whether the query set shares the same subgraph. Therefore, CSR is the only state-of-the-art method designed to address the I-FKGC task. However, due to the limited number of supporting triples, the extracted subgraph may not generalize well to the query set, which limits its performance.

\subsection{Neural Process Family}
Neural Process (NP) \cite{garnelo2018neural} attempts to define a distribution over prediction functions with limited observed data, which can quickly be adapted to new tasks. CNP \cite{garnelo2018conditional} encodes the observed data into a deterministic hidden variable, which fails to account for uncertainties. After that, there are several studies proposed to improve the neural process in various aspects. For instance, ANP \cite{kim2019attentive} introduces the self-attention mechanism to better capture dependencies and model the distribution. SNP \cite{singh2019sequential} is designed for sequential data and employs a recurrent neural network (RNN) to capture temporal correlation for better generalization. NP has been applied to many applications, such as recommender systems \cite{lin2021task} and link prediction \cite{liang2022neural}, \cite{luo2023graph}. Recently, NP-FKGC \cite{luo2023normalizing} proposes a normalizing flow-based NP for few-shot knowledge graph completion, and RawNP \cite{zhao2023towards} proposes a relational anonymous walk-based NP to inductively predict the missing facts for unseen entities. 

\subsection{Stochastic Attention Mechanism}
The stochastic attention mechanism is a novel attention mechanism based on a stochastic model that attempts to capture complicated dependencies and regularize the weights based on the input data. Early works \cite{bahuleyan2017variational,shankar2019posterior} adopt a normal distribution as the posterior distribution of the attention weight, satisfying the simplex constraints, \textit{i.e.,} sum to one \cite{bahuleyan2017variational}. However, these methods cannot utilize back-propagation to optimize the attention weight. Recently, the Bayesian attention module \cite{fan2020bayesian} proposes a differentiable stochastic attention, which can be optimized during training. The stochastic attention has been applied in many applications. Xu et al. \cite{xu2015show} adopt stochastic attention to capture the important regions in the image for image captioning. GSAT \cite{miao2022interpretable} extracts label-relevant subgraphs with graph stochastic attention to provide interpretable explanations for predictions.  
\section{Preliminaries}
\label{sec:pre}
In this section, we give a brief introduction to the key concepts in our paper and a formal definition of our problem. Commonly used notations are present in Table~\ref{tab:notations}.

\subsection{Neural Process}
Combining the benefits of the stochastic process and neural networks, neural process (NP) \cite{garnelo2018neural} attempts to model the joint distribution over the prediction function $f: X \rightarrow Y$ given limited data, where $X$ and $Y$ denote the input feature and label, respectively. The function $f$ is parameterized by a high-dimensional random vector $z$ whose distribution is denoted as $P(z|\mathcal{C})$. This distribution is conditioned on the context data $\mathcal{C}=\left\{\left(x_\mathcal{C}, y_\mathcal{C}\right)\right\}$ and estimated by an \textit{encoder}. For instance, if we assume that the distribution of $z$ follows a Gaussian distribution, the encoder would predict the mean and variance of this Gaussian distribution.

\begin{table}[t]
\centering
\caption{Glossary of commonly used notations.}
\label{tab:notations}
\resizebox{.9\columnwidth}{!}{%
    \begin{tabular}{cl}
        \toprule
        \midrule
        \textbf{Notation} & \textbf{Definition} \\
        \midrule
        $\mathcal{G}$ & knowledge graph \\ \midrule
        $\mathcal{E}, \mathcal{R}$ & the sets of entities and relations\\ \midrule
        $\mathcal{T}$ & a collection of triples \\ \midrule
        $h,r,t$ & head entity, relation, and tail entity \\ \midrule
        $\mathcal{C}_{r_q}$ & context data (\textit{a.k.a} support set) given a new relation $r_q$ \\  \midrule
        $\mathcal{C}_{r_q}^{-}$ & negative triples for context data \\  \midrule
        $\mathcal{D}_{r_q}$ & target data (\textit{a.k.a} query set) given a new relation $r_q$ \\  \midrule
        $\rho, z$ & hypotheses extracted from context data \\  \midrule
        $\theta, \phi, \psi$ & parameters of encoder, decoder, and extractor, respectively \\ \midrule
        $f_{\rho}(\cdot)$ & function to check whether the hypothesis exists \\  \midrule
        $\mathcal{G}_{(h,t)}$ & enclosing subgraph associated with $h$ and $r$ \\ \midrule
        $\mathcal{N}_{k}(h)$ & $k$-hop neighbor triples of $h$ \\ \midrule
        $\mathbbm{1}(\cdot)$ & Indicator function specifying whether two nodes are the same \\ \midrule
        $\|$ & concatenation function \\ \midrule
        $W, b, \sigma$ & learnable transformation matrix, bias, and nonlinear activation function \\ \midrule
        $\mathbf{h}_{\mathcal{G}_{(h,t)}}, \mathbf{h}_{\mathcal{G}_{S}}$ & representation of subgraph \\ \midrule
        $y_i$ & an indicator vector to denote whether the triple is positive or not \\ \midrule
        $\mathcal{N}(\mu(\mathbf{z}), \sigma(\mathbf{z}))$ & the Gaussian distribution parameterized by $\mathbf{z}$ \\ \midrule
        $\mathcal{G}_{S}$ & subgraph identified by the hypothesis \\ \midrule
        $M_z$ & soft edge mask \\ \midrule
        $\odot$ & element-wise product \\ \midrule
        $g_{\psi}(\cdot)$ & extractor to extract the subgraph $\mathcal{G}_{S}$ grounded by hypothesis $z$\\ \midrule
        $f_{\phi}(\cdot)$ & measurement for the plausibility of the triple \\  \midrule
        \bottomrule
    \end{tabular}%
}
\end{table}

NP can readily adapt to new prediction tasks by sampling a $z$ from the distribution. The prediction likelihood over \textit{target data} $\mathcal{D}=\left\{\left(x_\mathcal{D}, y_\mathcal{D}\right)\right\}$ is modeled as
\begin{equation}
\setlength\abovedisplayskip{1pt}
\setlength\belowdisplayskip{1pt}
\label{equ:equ1}
P\left(y_\mathcal{D}|x_\mathcal{D}, \mathcal{C}\right)=\int_z P\left(y_\mathcal{D}|x_\mathcal{D}, z\right) P(z|\mathcal{C})dz,
\end{equation}
where $P\left(y_\mathcal{D}|x_\mathcal{D}, z\right)$ is modeled by a \textit{decoder} network. Due to the intractable actual distribution of $z$, the optimization of the NP is achieved by amortized variational inference \cite{kingma2013auto}. The objective function in Eq.~\eqref{equ:equ1} can be optimized by maximizing the evidence lower bound (ELBO), which is formulated as
\begin{align}
    \label{equ:equ2}
    \log P\left(y_\mathcal{D}|x_\mathcal{D}, \mathcal{C}\right) 
    &\geq \mathbb{E}_{Q_\theta(z|\mathcal{C}, \mathcal{D})}\left[\log P_\phi\left(y_\mathcal{D}|x_\mathcal{D}, z\right)\right] \nonumber \\ 
    &-KL\left(Q_\theta(z|\mathcal{C}, \mathcal{D}) \| P_\theta(z|\mathcal{C})\right),
\end{align}
where $\theta$ and $\phi$ represent the parameters of \textit{encoder} and \textit{decoder}, respectively, and $Q_\theta(z|\mathcal{C}, \mathcal{D})$ denotes the estimation of the actual posterior distribution.

\subsection{Graph Stochastic Attention}
In graph learning tasks (\textit{e.g.,} graph classification), the goal is to predict the label $Y$ of a given graph $\mathcal{G}$ using a graph prediction function $f: \mathcal{G}\to Y$.
Inspired by the principle of information bottleneck \cite{tishby2015deep}, graph stochastic attention (GSAT) \cite{miao2022interpretable} aims to identify a subgraph $\mathcal{G}_S$ that mostly indicates the label $Y$ by imposing an information constraint. This can be formulated as
\begin{equation}
\label{equ:gib}
\min\nolimits_{\psi}-I\left(\mathcal{G}_S ; Y\right)+\beta\cdot I\left(\mathcal{G}_S; \mathcal{G}\right) \text {, s.t. } \mathcal{G}_S \sim g_{\psi}(\mathcal{G}),
\end{equation}
where $g_{\psi}(\mathcal{G})$ denotes the extractor with parameter $\psi$ to extract possible subgraphs from $\mathcal{G}$, $I(\cdot, \cdot)$ indicates the mutual information (MI) between two random variables, and $\beta$ controls the trade-off between the maximization of $I\left(\mathcal{G}_S; Y\right)$ and the minimization of $I\left(\mathcal{G}_S; \mathcal{G}\right)$.
By optimizing Eq.~\eqref{equ:gib}, we encourage the $\mathcal{G}_S$ to inherit the relevant information in $Y$, while minimizing the irrelevant information from the input $\mathcal{G}$. This allows $\mathcal{G}_S$ to effectively reveal patterns for predictions and demonstrate explainability.

Following previous works \cite{alemi2016deep}, we can derive a tractable variational upper bound of Eq.~\eqref{equ:gib}, which is formulated as
\begin{equation}
\resizebox{\columnwidth}{!}{$
\begin{aligned}
    \min _{\psi, \phi} -\mathbb{E}_{\mathcal{G}_S, Y}\left[\log P_\phi\left(Y | \mathcal{G}_S\right)\right]&+\beta\cdot \mathbb{E}_\mathcal{G}\left[\operatorname{KL}(P_{\psi}(\mathcal{G}_S | \mathcal{G}) \| Q (\mathcal{G}_S))\right], \\
    \text { s.t. } \mathcal{G}_S &\sim P_{\psi}(\mathcal{G}_S | \mathcal{G}),
\end{aligned}
$}
\label{equ:gsat}
\end{equation}
where $P_{\psi}\left(\mathcal{G}_S | \mathcal{G}\right)$ denotes an \textit{extractor} with parameter $\psi$ that extracts the subgraph $\mathcal{G}_S$ from $\mathcal{G}$, and $P_\phi\left(Y | \mathcal{G}_S\right)$ denotes a \textit{predictor} with parameter $\phi$ that predicts the label $Y$ based on the subgraph $\mathcal{G}_S$. The $Q \left(\mathcal{G}_S\right)$ is the variational approximation for the marginal distribution of $\mathcal{G}_S$.

\subsection{Problem Definition}
A knowledge graph (KG) can be denoted as $\mathcal{G}=\{\mathcal{E}, \mathcal{R}, \mathcal{T}\}$, where $\mathcal{E}$ and $\mathcal{R}$ respectively represent the set of entities and relations. The $\mathcal{T}=\{(h, r, t) \subseteq \mathcal{E} \times \mathcal{R} \times \mathcal{E}\}$ represents a collection of triples, where $h$, $r$, and $t$ denote the head entity, relation, and tail entity, respectively. 

Conventional few-shot KG completion (FKGC) aims to predict missing facts for a new relation in the \textit{query set} with a few-shot triple given in the \textit{support set}. Previously, entities in the support set and query set are assumed to exist in the KGs. However, in real-world scenarios, new relations and entities may emerge simultaneously that are not included in the original background KGs. This paper focuses on this more challenging task, denoted as inductive few-shot KG completion (I-FKGC).

\vspace{3mm} \textbf{Inductive Few-shot Knowledge Graph Completion (I-FKGC).}
Given a new relation $r_q\notin\mathcal{R}$, and its associated $K$-shot support set $\left\{\left(h_i, r_q, t_i\right)\right\}_{i=1}^K$, we aim to predict the other entity $t_q$ for each query in the query set $\left\{\left(h_q, r_q, t_q\right)\right\}$, where all the entities in the query set do not exist in the background knowledge graph, \textit{i.e.,} $h_q,t_q \notin \mathcal{E}$.

In our paper, we propose a neural process-based framework for the I-FKGC task. Formally, given a new relation $r_q$, we treat its support set as the context data $\mathcal{C}_{r_q}=\left\{\left(h_i, r_q, t_i\right) \right\}_{i=1}^K$ and the query set as the target data $\mathcal{D}_{r_q}=\left\{\left(h_q, r_q, ?\right)\right\}$.


\section{Approach}
\label{sec:approach}
In this section, we first discuss the motivation from the inductive reasoning perspective, and then we introduce the proposed method.

\begin{figure*}[ht]
    \centering
    \includegraphics[width=.9\linewidth]{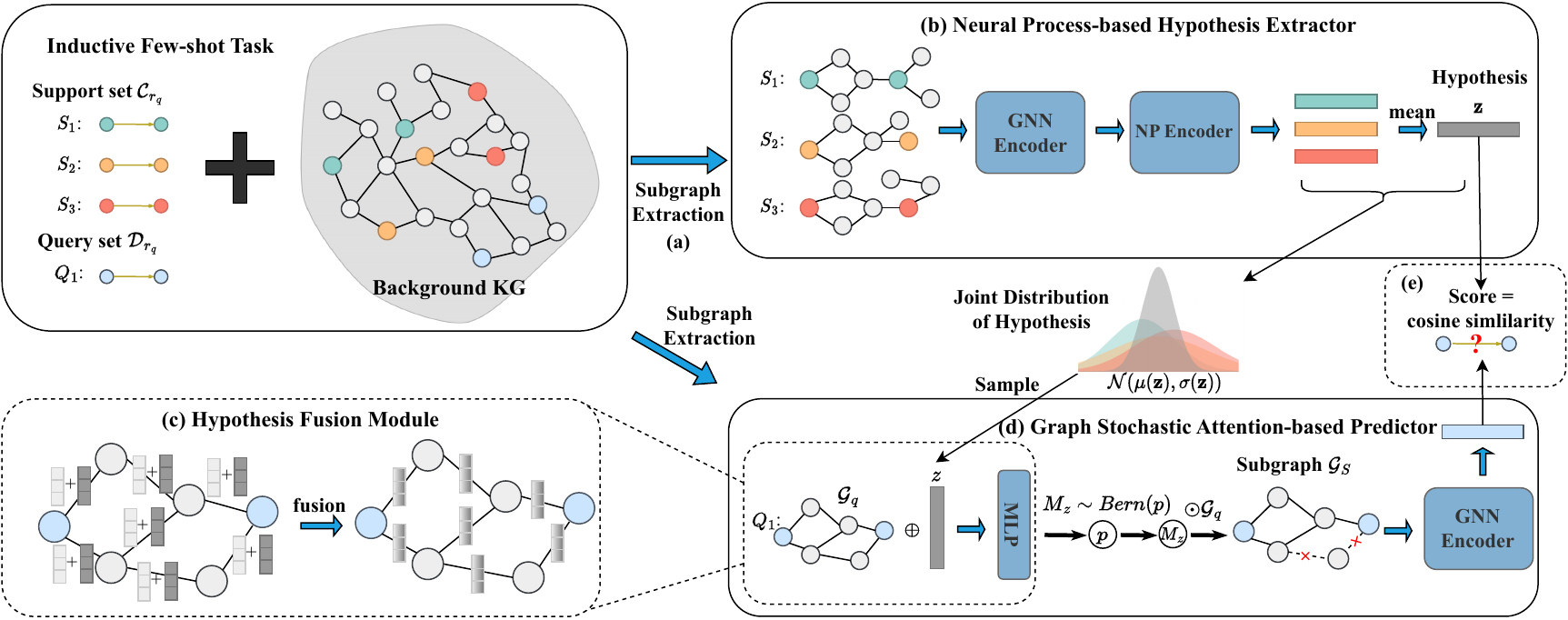}
    \caption{The overall framework of our proposed model \ourmethod, which consists of two major modules, namely, a neural process hypothesis extractor (including (b)) and a graph stochastic attention-based predictor (including (c), (d), and (e)). (a) We first extract enclosing subgraphs for all triples from the background knowledge graph. (b) We adopt the GNN encoder and NP encoder to model the joint distribution of the hypothesis. (c) After sampling a hypothesis (\textit{i.e.}, $z$) from the distribution, we inject the hypothesis into the graph structure by the hypothesis fusion module. (d) We apply the graph stochastic attention to identify a subgraph and feed it into the GNN encoder to get representation. (e) We compute the cosine similarity between the subgraph representation and the hypothesis to test whether the query aligns with the shared hypothesis extracted from the support set.}
    \label{fig:architecture}
\end{figure*}

\subsection{I-FKGC as an Inductive Reasoning Problem}

Inductive reasoning \cite{DanHunter} is the process of formulating a general hypothesis based on past observations and applying it to predict future events. Inspired by this, we cast the inductive few-shot link prediction as an inductive reasoning problem. Specifically, we try to find a hypothesis (\textit{i.e.,} shared subgraph of all triples) in the support set and then test whether the hypothesis can be applied to the query set for predictions.

\vspace{3mm} \textbf{Example.} As the example shown in \Cref{fig:introduction}(b), given a new relation $r_q$, we want to find a hypothesis that could explain how $h$ establishes a relation $r_q$ with the $t$ for prediction. By examining the subgraph structure surrounding the triples, we can observe a shared connection pattern between $h_k$ and $t_k$ in the support set from background KGs $\mathcal{G}$. 
In this case, we can extract a hypothesis $\rho$ from the support set as:
\begin{equation}
    \rho := (h_q, \texttt{requires}, e_q) \wedge (e_q, \texttt{works in}, t_q) \to (h_q, r_q, t_q), \nonumber
\end{equation}
which means that if the hypothesis on the left-hand side is true, we can deduce the target relation between entities as $(h_q, r_q, t_q)$. Thereby, the prediction function based on the hypothesis for I-FKGC can be formulated as $f_\rho: \rho(h_q,r_q,t_q)\to y, y\in\{0,1\}$.


Based on the above motivation, we design the overall framework of \ourmethod, as illustrated in \Cref{fig:architecture}, for the I-FKGC task, which consists of two major modules, namely: 

\begin{itemize}
    \item \textbf{A neural process-based hypothesis extractor} to generate the hypothesis shared in the support set. To this end, we employ GNN and Neural Process (NP) encoders to capture the structure dependence of entities and model the joint distribution of the potential hypothesis, based on which we obtain a hypothesis to capture the shared pattern in the support set.
    \item \textbf{A graph stochastic attention-based predictor} to test whether the hypothesis can be applied to the query set for predictions. By injecting the extracted hypothesis into the graph structure, we develop a graph stochastic attention module to identify a subgraph from the query triple, whose representation will be compared with the extracted hypothesis, to determine the plausibility of triples.
\end{itemize}

We provide details on these two components below.



\subsection{Neural Process-based Hypothesis Extractor}
Neural process-based hypothesis extractor attempts to extract the inductive hypothesis from the support set for predictions. Existing methods extract the hypothesis by finding the most commonly shared subgraph structure among the support set \cite{huang2022few}. However, due to the limited number of support triples, the extracted hypothesis may not generalize effectively to the query set. Drawing inspiration from the neural process, we propose to model the joint distribution of hypothesis. This allows us to sample a hypothesis from the distribution that generalizes well to the query set instead of overfitting to the sparse data.
%


\subsubsection{Graph Neural Network Encoder}
The graph structure inherently encodes the patterns for inductive reasoning \cite{teru2020inductive}. Therefore, we extract the enclosing subgraph $\mathcal{G}_{(h, t)} \subset \mathcal{G}$ for each triple $\left(h, r_q, t\right)$ in the support set, which can be formulated as
\begin{equation}
    \label{eq_contex}
    \mathcal{G}_{(h, t)} = \mathcal{N}_k(h) \cap \mathcal{N}_k(t),
\end{equation}
where $\mathcal{N}_k(h)$, $\mathcal{N}_k(t)$ denotes the $k$-hop neighbor triples of $h$, $t$, respectively. The value of $k$ is set to 1 or 2 according to the density of KGs \cite{huang2022few}. The enclosing subgraph captures the relevant structural information about the entities $h$ and $t$, which can be used to deduce the potential relation.

To extract the shared features of all the subgraphs in the support set, we adopt a GNN encoder called PathCon \cite{wang2021relational} to obtain the representation of each subgraph $\mathbf{h}_{\mathcal{G}_{\left(h,t\right)}}$. The PathCon focuses solely on the edge feature $e_r$ through a relation message passing scheme, which enhances its ability to capture structural patterns for hypothesis extraction. This is formulated as
\begin{align}
    a_v^l& = \sum_{(v,r,u) \in \mathcal{N}_1(v)} e_{r}^l, \label{equ:rms1}  \\
    e_v^l& = a_v^l\|\mathbbm{1}(v=h)\| \mathbbm{1}(v=t), \\
    e_r^{l+1} & =\sigma\left(\left(e_v^l \| e_u^l \| e_r^l\right) \cdot W^l+b^l\right), (v, r, u) \in \mathcal{G}_{(h,t)}, \label{equ:rms3}
\end{align}
where ``$\|$'' is the concatenation function, $a_v^l$ is the representation of entity $v$ at layer $l$, $(v,r,u) \in \mathcal{N}_1(v)$ denotes all the triples directly connecting to $v$, $e_{r}^l$ is the hidden representation of edge $r$. The $e^1_r$ can be initialized randomly or from pretrained relation embedding and $e_v^l$ denotes the hidden representation of entity $v$. The operation $\mathbbm{1}(\cdot)$ indicates whether two nodes are the same and returns 1 if same and 0 for different, which injects the information of target entities $h,t$ into the subgraph representation. $W^l$, $b^l$, and $\sigma$ denote the learnable transformation matrix, bias, and nonlinear activation function.


By stacking $L$ layers, we can obtain the subgraph representation $\mathbf{h}_{\mathcal{G}_{\left(h,t\right)}}$ by summarizing all the edge representations with max pooling and concatenating them with target entities representations as
\begin{align}
     & \mathbf{e}  = \text{MaxPooling}(e_r^L|r\in\mathcal{G}_{(h,t)}),             \\
     & \mathbf{h}_{\mathcal{G}_{\left(h,t\right)}} = \mathbf{e} \| a_h^L \| a_t^L.  \label{equ:subgraph}
\end{align}

\subsubsection{Neural Process Encoder}
The $\mathbf{h}_{\mathcal{G}_{\left(h,t\right)}}$ encodes the inductive features that can be used to deduce the relation. Therefore, the neural process encoder tries to capture connections between the hidden representation and the target relation from the support set to model the distribution of the hypothesis $P(\mathbf{z} | \mathcal{C}_{r_q})$.

For each triple $(h_i,r_q,t_i)$ in the support set, we first get its subgraph representation $\mathbf{h}_{\mathcal{G}_{\left(h_i,t_i\right)}}$ by GNN encoder. To reduce the estimation bias of the distribution, we generate a few negative triples in the support set as $\mathcal{C}^{-}_{r_q}$ by corrupting the head or tail entities \cite{10181235}. Then, we obtain their subgraph representations with the same GNN encoder.

To capture the connections for representing the hypothesis, we concatenate the subgraph representation with an indicator vector $y_i$ to generate $c_i$ as follows:
\begin{equation}
    c_i = \operatorname{MLP}\left(\mathbf{h}_{\mathcal{G}_{\left(h_i,t_i\right)}} \| y_i\right),  y_i=\left\{\begin{array}{l}
        1,\left(h_i, r_q, t_i\right) \in \mathcal{C}_{r_q} \\
        0,\left(h_i, r_q, t_i\right) \in \mathcal{C}_{r_q}^{-}
    \end{array}\right.,
    \label{equ:c}
\end{equation}
where $y_i$ denotes whether the triple is positive or not.

Then, we summarize all the latent representations $c_i \in \mathcal{C}_{r_{q}} \cup \mathcal{C}_{r_q}^{-}$ into a global representation $\mathbf{z}$ by a \textit{permutation-invariant} aggregator function \cite{garnelo2018neural, van1976stochastic}, to model the joint distribution over hypothesis. In our paper, we select a simple average function, which is formulated as
\begin{equation}
    \mathbf{z}=\frac{1}{\left|\mathcal{C}_{r_q} \cup \mathcal{C}_{r_q}^{-}\right|} \sum_{c_i \in \mathcal{C}_{r_q} \cup \mathcal{C}_{r_q}^{-}} c_i.
\end{equation}

The distribution $P(\mathbf{z} | \mathcal{C}_{r_q})$ is empirically assumed to follow the Gaussian distribution $\mathcal{N}(\mu(\mathbf{z}), \sigma(\mathbf{z}))$, which is parameterized by $\textbf{z}$. The mean $\mu(\mathbf{z})$ and variance $\sigma(\mathbf{z})$ can be estimated by using two neural networks as follows
\begin{gather}
    \chi=\operatorname{ReLU}(\operatorname{MLP}(\mathbf{z})), \\
    \mu(\mathbf{z})=\operatorname{MLP}\left(\chi\right), \\
    \sigma(\mathbf{z})=0.1+0.9 \cdot \operatorname{Sigmoid}\left(\operatorname{MLP}\left(\chi\right)\right). \label{equ:sigmaz}
\end{gather}

Finally, we can sample a $z$ from the distribution as the instance of the hypothesis, which is formulated as
\begin{equation}
    \text { Sample } z \sim \mathcal{N}(\mu(\mathbf{z}), \sigma(\mathbf{z})), \label{equ:sample}
\end{equation}
where $z$ is considered as the hidden representation of the hypothesis, which will be used for predictions.

\subsection{Graph Stochastic Attention-based Predictor}
Graph stochastic attention-based predictor is designed to test the extracted hypothesis on the query for predictions. However, as the extracted hypothesis $z$ is a latent variable, it cannot be directly applied to the graph structure. To solve this problem, we employ the graph stochastic attention mechanism \cite{fan2020bayesian} to transform the hypothesis $z$ into a soft edge mask $M_z$. This mask allows us to extract a subgraph $\mathcal{G}_S$ identified by the hypothesis, which can then be used for prediction and offering explanations.


For each triple in the query set $\mathcal{D}_{r_q}$, we first extract its enclosing subgraph $\mathcal{G}_{(h, t)}$ by Eq.~\eqref{eq_contex}.
%
Following the paradigm of graph stochastic attention, we attempt to learn an \textit{extractor} $g_{\psi}$ to model the distribution $P_{\psi}(\mathcal{G}_S|\mathcal{G}_{(h,t)},z)$ and extract a subgraph $\mathcal{G}_S$ grounded by hypothesis $z$. Specifically, we assume the distribution $P_{\psi}(\mathcal{G}_S|\mathcal{G}_{(h,t)},z)$ following the Bernoulli distribution \cite{miao2022interpretable}, which is formulated by the joint probability of all edges in the subgraph $\mathcal{G}_S$ as follows
\begin{gather}
    P_{\psi}(\mathcal{G}_S|\mathcal{G}_{(h,t)},z) = \prod_{(h,r,t) \in {\mathcal{G}_{(h, t)}}} P_{\psi}(r|e_r,z), \label{equ:psi}\\
    P_{\psi}(r|e_r, z) = \text{Sigmoid}(\text{MLP}_{\psi}(e_r + z)),
\end{gather}
where $P_{\psi}(r|e_r,z) \in [0,1]$ denotes the existence possibility of each edge in $\mathcal{G}_S$, which is estimated by the \emph{hypothesis fusion module}, and ``$+$'' denotes the element-wise addition. $\text{MLP}_{\psi}$ and $\text{Sigmoid}$ denote the multi-layer perception with parameter $\psi$ and an activation function, respectively.

By sampling from $P_{\psi}(\mathcal{G}_S|\mathcal{G}_{(h,t)},z)$, we can obtain an edge mask $M_z$ to extract the subgraph as
\begin{gather}
    \text{Sample}~M_z \sim \operatorname{Bern}(P_{\psi}(\mathcal{G}_S|\mathcal{G}_{(h,t)},z)), \\
    \mathcal{G}_S = \mathcal{G}_{(h, t)} \odot M_z,  \label{equ:gs}
\end{gather}
where ``$\odot$'' denotes the element-wise product between edges and mask. To ensure the computable gradient of $P_{\psi}(r|e_r, z)$, we apply the gumbel-softmax reparameterization \cite{jang2016categorical} to sample mask $M_z$.




The extracted subgraph $\mathcal{G}_S$ is then fed into the \textit{predictor} $f_\phi$ to test whether the query set aligns with the hypothesis. We first adopt the same GNN encoder in the hypothesis extractor to obtain the subgraph representation $\mathbf{h}_{\mathcal{G}_S}$. Then we compute the similarity (\textit{e.g.,} cosine similarity) between the subgraph representation $\mathbf{h}_{\mathcal{G}_S}$ and the hypothesis $z$ to measure the plausibility of the triple, which can be formulated as
\begin{equation}
    \label{eq:gsat}
    f_\phi(\mathcal{G}_S, z) = \text{Cosine}(\mathbf{h}_{\mathcal{G}_S}, z).
\end{equation}




\begin{algorithm2e}[t]
    \caption{The training process of \ourmethod}\label{alg:gs_np}
    \KwIn {Knowledge graph $\mathcal{G}$; Training relations $\mathcal{R}_{train}$}
    \KwOut {encoder parameters: $\theta$, decoder parameters: $\phi$, extractor parameters: $\psi$}
    Initialization: $\theta$, $\phi$, and $\psi$ randomly\;
    \While{not converge}{
        Sample a relation $\mathcal{T}_{r_q} = \{\mathcal{C}_{r_q}, \mathcal{D}_{r_q}\}$ from $\mathcal{R}_{train}$\;
        Extract the enclosing subgraph $\mathcal{G}_{(h,t)}$ for each triple $(h,r_q,t) \in \mathcal{T}_{r_q}$ (Eq.~\eqref{eq_contex})\;
        Generate the subgraph representation $\mathbf{h}_{\mathcal{G}_{(h,t)}}$ (Eqs.~\eqref{equ:rms1}-\eqref{equ:subgraph})\;
        Generate the prior distribution $P(z|\mathcal{C}_{r_q})$ by $\mathcal{C}_{r_q}$ (Eqs.~\eqref{equ:c}-\eqref{equ:sigmaz})\;
        Generate the variational posterior distribution $Q_{\theta}\left(z|\mathcal{C}_{r_q}, \mathcal{D}_{r_q}\right)$ by $\mathcal{C}_{r_q}, \mathcal{D}_{r_q}$ (Eqs.~\eqref{equ:c}-\eqref{equ:sigmaz})\;
        Sample a $z$ from the posterior distribution $Q_{\theta}\left(z|\mathcal{C}_{r_q}, \mathcal{D}_{r_q}\right)$ (Eq.~\eqref{equ:sample})\;
        Extract a subgraph $\mathcal{G}_S$ given the $z$, $\mathcal{G}_{(h,t)}$ (Eqs.~\eqref{equ:psi}-\eqref{equ:gs})\;
        Calculate the similarity between $\mathcal{G}_S$ and $z$ (Eq.~\eqref{eq:gsat})\;
        Optimize $\theta$, $\phi$, and $\psi$ using ELBO loss (Eq.~\eqref{equ:loss})\;
    }
\end{algorithm2e} 
\subsection{Optimization}
\ourmethod draws inspiration from both neural processes and graph stochastic attention. In this section, we present the optimization process of \ourmethod, which seamlessly combines the optimization of neural processes and graph stochastic attention into a unified objective function. The training process of \ourmethod is illustrated in Algorithm~\ref{alg:gs_np}.

Given a relation $r_q$ and corresponding support set and query set, \ourmethod aims to extract a hypothesis $z$ from the support set $\mathcal{C}_{r_q}$ and minimize the prediction loss on the query set $\mathcal{D}_{r_q}$. This can be optimized by maximizing the evidence lower bound (ELBO), which can be derived as
\begin{equation}
    \resizebox{\columnwidth}{!}{$
    \begin{aligned}
         & \log P\left(t_q | h_q, r_q, \mathcal{C}_{r_q},\mathcal{G}_q\right)
        \geq \int_Z Q(Z) \log \frac{P(t_q, Z | h_q, r_q, \mathcal{C}_{r_q},\mathcal{G}_q)}{Q(Z)} \\
         & =\mathbb{E}_{Q(Z)}\left[\log P(t_q | h_q, r_q, Z)+\log \frac{P(\mathcal{G}_S,z | h_q, r_q, \mathcal{C}_{r_q}, \mathcal{G}_q)}{Q(Z)}\right]        \\
         & =\underbrace{\mathbb{E}_{Q(Z)}\left[\log P\left(t_q | h_q, r_q, Z\right)\right]}_{\text{(1)}}-\underbrace{K L(Q(z) \| P(z | \mathcal{C}_{r_q}))}_{\text{(2)}} \\
         & -\underbrace{K L(Q(\mathcal{G}_S) \| P(\mathcal{G}_S | \mathcal{G}_q, z))}_{\text{(3)}},
        \label{eq:objective}
    \end{aligned}$}
\end{equation}
where $Z=(\mathcal{G}_S, z)$ denotes all the latent variables, and $\mathcal{G}_q$ denotes the original enclosing subgraph of the query $\mathcal{G}_{(h_q,t_q)}$ for simplicity.
The hypothesis $z$ depends on the support set $\mathcal{C}_{r_q}$ and extracted subgraph $\mathcal{G}_S$ depends on the query set $\mathcal{D}_{r_q}$, support set $\mathcal{C}_{r_q}$, and hypothesis $z$. The $Q(z)$ is the true posterior distribution of $z$, which is approximated by $Q_\theta(z | \mathcal{C}_{r_q}, \mathcal{D}_{r_q})$ during training. The $Q(\mathcal{G}_S)$ denotes the variational approximation to the true distribution over the subgraph. The true posterior distribution $z$ and $\mathcal{G}_S$ (\textit{a.k.a.} $Q(z)$ and $Q(\mathcal{G}_S)$) are independent from each other. 
The derivation of the ELBO loss in Eq.~\eqref{eq:objective} can be found in the Appendix~\ref{sec:elbo}.

From Eq.~\eqref{eq:objective}, we can see that the combination of part (1) and part (2) is identical to the objective of the neural process, while the combination of part (1) and part (3) is the objective of graph stochastic attention. Therefore, the Eq.~\eqref{eq:objective} seamlessly combined the objectives of these two modules which are jointly optimized during training. 

According to the GSAT \cite{miao2022interpretable}, the third KL-divergence term in Eq.~\eqref{eq:objective} can be further computed as
\begin{equation}
    \label{eq:r}
    \begin{aligned}
         & \operatorname{KL}\left(Q\left(\mathcal{G}_S\right) \| P\left(\mathcal{G}_S | \mathcal{G}_q, z\right) \right)=                                 \\
         & \sum_{(h,r,t) \in \mathcal{G}_q} p_{e_{r}} \log \frac{p_{e_{r}}}{\tau}+\left(1-p_{e_{r}}\right) \log \frac{1-p_{e_{r}}}{1-\tau}+c(n, \tau),
    \end{aligned}
\end{equation}
where $p_{e_{r}}$ is the probability predicted by hypothesis fusion module, $n$ denotes the number of edges in $\mathcal{G}_q$, $\tau \in [0,1]$ is a hyper-parameter, and $c(n,\tau)$ is a constant value.

The final objective function is presented as follows:
\begin{equation}
\resizebox{\columnwidth}{!}{$
    \begin{aligned}
        \label{equ:target}
         & \mathcal{L}(\phi, \theta) = \mathbb{E}_{Q(\mathcal{G}_S, z)}\left[\log P_{\phi}\left(t_q | h_q, r_q, \mathcal{G}_S, z\right)\right] \\
         & -\operatorname{KL}(Q_\theta(z | \mathcal{C}_{r_q}, \mathcal{D}_{r_q}) \| P_\theta(z | \mathcal{C}_{r_q})) - \operatorname{KL}(Q(\mathcal{G}_S) \| P(\mathcal{G}_S | \mathcal{G}, z)),
    \end{aligned}$}
\end{equation}
where $\theta$ and $\phi$ denote the parameters of \textit{encoder} and \textit{predictor} (\textit{a.k.a.,} \textit{decoder}).

To support gradient propagation, we introduce the \textit{reparamterization trick} for sampling $z$, and then we estimate the expectation $\mathbb{E}_{Q(\mathcal{G}_S, z)}\left[\log P_{\phi}\left(t_q | h_q, r_q, \mathcal{G}_S, z\right)\right]$ via the Monte-Carlo sampling as

\begin{equation}
\resizebox{\columnwidth}{!}{$
    \begin{aligned}
        \mathbb{E}_{Q(\mathcal{G}_S, z)}\left[\log P_{\phi}\left(t_q | h_q, r_q, \mathcal{G}_S, z\right)\right] & \simeq \frac{1}{T} \sum_{t=1}^T \log P_{\phi}\left(t_q | h_q, r_q, \mathcal{G}_S, z^{(t)}\right), \\ 
    z^{(t)}=\mu(\mathbf{z})+\sigma(\mathbf{z}) \epsilon^{(t)}, & \text { with } \epsilon^{(t)} \sim \mathcal{N}(0,1).
    \end{aligned}$}
\end{equation}

The first term in Eq.~\eqref{equ:target} represents the prediction likelihood of both the decoder in NPs and the predictor in graph stochastic attention. This can be calculated using the commonly used margin ranking loss, which is
\begin{equation}
    \log P_{\phi}(t_q | h_q, r_q, \mathcal{G}_S, z)=-\sum_{q, q^{-}} \max (0, \gamma+s(q^{-})-s(q)), \label{equ:loss}
\end{equation}
where $\gamma$ denotes a margin hyper-parameter, and $q$, $q^-$ represent true triples and negative triples, respectively. We attempt to rank the scores of true triples higher than all other negative triples by maximizing the likelihood.

\subsection{Complexity Analysis}
In this section, we will concisely outline the complexity analysis of our method. Given a relation $r_q$ with its $K$-shot support set $\mathcal{C}_{r_q}$ and $m$ factual triples in the query set $\mathcal{D}_{r_q}$, the complexity of \ourmethod is $O((n+1)K + m)$, where $n$ denotes the negative sampling size for each triple in $\mathcal{C}_{r_q}$. Our method only needs to encode each triple in $\mathcal{C}_{r_q} \cup \mathcal{C}_{r_q}^{-}$ and predict triples in the query set.

\section{Experiments}
\label{sec:exp}
In this section, we will qualitatively and quantitatively demonstrate the superiority of our method.
\subsection{Datasets and Evaluation}
\begin{table*}[ht]
        \centering
        \caption{Statistics of datasets. Ind-BG refers to the background KG used in training time and Ind-Test refers to the knowledge graph constructed by all entities and triples involved in the test tasks.}
        \label{tab:dataset}
        \resizebox{.9\textwidth}{!}{%
                \begin{tabular}{@{}ccccc|cccc|cccc@{}}
                        \toprule
                        \multirow{2}{*}{KG} & \multicolumn{4}{c|}{NELL} & \multicolumn{4}{c|}{ConceptNet} & \multicolumn{4}{c}{WIKI}                                                                                                   \\ \cmidrule(l){2-13}
                                            & \#rels                    & \#entities                      & \#edges                  & \#tasks & \#rels & \#entities & \#edges   & \#tasks & \#rels & \#entities & \#edges   & \#tasks \\ \midrule
                        Ind-BG              & 291                       & 44,005                          & 82,318                   & -       & 14     & 619,163    & 1,191,782 & -       & 822    & 2,583,905  & 3,221,617 & -       \\
                        Ind-Test            & 291                       & 24,539                          & 98,791                   & 11      & 14     & 171,540    & 1,350,214 & 2       & 822    & 2,179,254  & 2,637,623 & 33      \\ \bottomrule
                \end{tabular}%
        }
\end{table*}

\begin{table*}[t]
        \centering
        \caption{The results of 3-shot I-FKGC on NELL, ConceptNet, and WIKI datasets.}
        \label{tab:main-table}
        \resizebox{.9\textwidth}{!}{%
                \begin{tabular}{p{2cm}|cccc|cccc|cccc}
                        \toprule
                             \multirow{2}{*}{Method}      & \multicolumn{4}{c|}{\textbf{NELL}} & \multicolumn{4}{c|}{\textbf{ConceptNet}} & \multicolumn{4}{c}{\textbf{WIKI}}                                                                                                                                                                                                                                                                                                                                              \\ \cmidrule(r){2-5} \cmidrule(r){6-9} \cmidrule(r){10-13}
                                   & \multicolumn{1}{c}{\textbf{MRR}}  & \multicolumn{1}{c}{\textbf{Hit@1}}      & \multicolumn{1}{c}{\textbf{Hit@5}} & \multicolumn{1}{c|}{\textbf{Hit@10}} & \multicolumn{1}{c}{\textbf{MRR}} & \multicolumn{1}{c}{\textbf{Hit@1}} & \multicolumn{1}{c}{\textbf{Hit@5}} & \multicolumn{1}{c|}{\textbf{Hit@10}} & \multicolumn{1}{c}{\textbf{MRR}} & \multicolumn{1}{c}{\textbf{Hit@1}} & \multicolumn{1}{c}{\textbf{Hit@5}} & \multicolumn{1}{c}{\textbf{Hit@10}} \\ \midrule

                        PathCon    & 0.003                             & 0.000                                   & 0.000                              & 0.000                               & 0.063                            & 0.000                              & 0.000                              & 0.000                               & 0.003                            & 0.000                              & 0.000                              & 0.000                               \\
                        GraIL      & 0.077                             & 0.049                                   & 0.070                              & 0.070                               & 0.046                            & 0.009                              & 0.016                              & 0.016                               & 0.026                            & 0.000                              & 0.000                              & 0.000                               \\
                        SNRI       & 0.195                             & 0.107                                   & 0.247                              & 0.291                               & 0.193                            & 0.138                              & 0.190                              & 0.231                               & 0.003                            & 0.000                              & 0.000                              & 0.000                               \\ \midrule
                        GMatching  & 0.288                             & 0.152                                   & 0.427                              & 0.601                               & 0.283                            & 0.151                              & 0.438                              & 0.548                               & 0.151                            & 0.059                              & 0.204                              & 0.340
                        \\
                        MetaR      & 0.251                             & 0.135                                   & 0.388                              & 0.494                               & 0.269                            & 0.151                              & 0.411                              & 0.493                               & 0.110                            & 0.038                              & 0.137                              & 0.244
                        \\
                        FSRL       & 0.204                             & 0.124                                   & 0.208                              & 0.236                               & 0.285                            & 0.164                              & 0.370                              & 0.493                               & 0.235                            & 0.163                              & 0.271                              & 0.326
                        \\
                        FAAN       & 0.485                             & 0.326                                   & 0.691                              & 0.803                               & 0.210                            & 0.096                              & 0.274                              & 0.438                               & 0.123                            & 0.076                              & 0.122                              & 0.145
                        \\
                        GANA       & 0.297                             & 0.163                                   & 0.461                              & 0.573                               & 0.239                            & 0.137                              & 0.329                              & 0.397                               & 0.118                            & 0.046                              & 0.115                              & 0.221
                        \\
                        NP-FKGC    & 0.354                             & 0.152                                   & 0.612                              & 0.787                               & 0.169                            & 0.014                              & 0.274                              & 0.630                               & 0.613                            & 0.550                              &      0.673                              &   0.712                                  \\
                        CSR        & \underline{0.511}                 & \underline{0.348}                       & \underline{0.725}                  & \underline{0.837}                   & \underline{0.611}                & \underline{0.496}                  & \textbf{0.729}                     & \textbf{0.786}                      & \underline{0.666}                & \underline{0.589}                  & \underline{0.744}                  & \underline{0.798}                   \\ \midrule
                        \ourmethod & \textbf{0.627}                    & \textbf{0.522}                          & \textbf{0.742}                     & \textbf{0.865}                      & \textbf{0.634}                   & \textbf{0.548}                     & \underline{0.712}                  & \underline{0.740}                   & \textbf{0.694}                   & \textbf{0.597}                     & \textbf{0.822}                     & \textbf{0.860}

                        \\ \bottomrule
                \end{tabular}%
        }
\end{table*}

We conduct evaluations on three widely used FKGC benchmarks, namely, NELL \cite{mitchell2018never}, ConceptNet \cite{speer2017conceptnet}, and WIKI \cite{vrandevcic2014wikidata}. Following the setting of CSR \cite{huang2022few}, we construct the datasets in an \textit{inductive manner}. We remove all entities in the test tasks, together with their one-hop neighbors, from the original KGs to construct an inductive background KG (\emph{Ind-BG}) during training. All entities and triples involved in the test tasks become the \emph{Ind-Test}. In the test time, we combine \emph{Ind-Test} with \emph{Ind-BG} to construct a test time background KG. 

Specifically, for the inductive setting, we mostly use the meta-eval and meta-test splits of NELL-One for the eval and test few-shot tasks on NELL and select the fewest 1/2 appearing relations as eval/test few-shot tasks for the ConceptNet, following the previous paper \cite{xiong2018one, lv2019adapting}. For each test task, we also subsample the number of query triplets to 10\%. 
Following the CSR, for the WIKI dataset, we first remove all entities and their one-hop neighbors involved in test tasks from the original knowledge graph and then subsample the number of query triples to $2\%$ to make sure that the remaining training time background KG does not become too small.
The statistics of datasets are summarized in Table \ref{tab:dataset}.

In the test time, we sample 50 negative tail candidates for each query triple and rank them together with the true tail entity. We adopt two widely used metrics, namely mean reciprocal rank (MRR) and the Top-$N$ hit ratio (Hit@$N$). Hit@$N$ measures the percentage of times that the positive tail is ranked higher than $N$ among the negative tail candidates. The $N$ is set to 1, 5, and 10 to fairly compare with existing methods \cite{huang2022few}.

\subsection{Baseline Models}
We compare the proposed method with two groups of approaches: \textbf{FKGC Methods}, including GMatching \cite{xiong2018one}, MetaR \cite{chen2019meta}, FSRL \cite{zhang2020few}, FAAN \cite{sheng2020adaptive}, GANA \cite{niu2021relational}, and NP-FKGC \cite{luo2023normalizing}; and \textbf{Inductive KGC Methods}, which include PathCon \cite{wang2021relational}, SNRI \cite{xu2022subgraph}, GraIL \cite{teru2020inductive}, and CSR \cite{huang2022few}. CSR is the current state-of-the-art (SOTA) method designed for the I-FKGC task.
The details of baseline models are shown as follows.

\textbf{FKGC methods.} This group of methods is all under the meta-learning framework, which can predict potential links given few-shot associated triples.
\begin{itemize}
    \item GMatching\footnote{\url{https://github.com/xwhan/One-shot-Relational-Learning}} \cite{xiong2018one} utilizes the entity embeddings and entity local graph structures to represent entity pairs and learn a metric to measure the similarity, which is the first work for FKGC.
    \item FAAN\footnote{\url{https://github.com/JiaweiSheng/FAAN}} \cite{sheng2020adaptive} adopts an adaptive neighbor encoder to encode entities and a Transformer to encoder entity pairs.
    \item MetaR\footnote{\url{https://github.com/AnselCmy/MetaR}} \cite{chen2019meta} attempts to learn relation-specific meta information from support triples and applies it to few-shot relational predictions.
    \item GANA\footnote{\url{https://github.com/ngl567/GANA-FewShotKGC}} \cite{niu2021relational} introduces a gated and attentive neighbor aggregator to represent entity representations.
    \item NP-FKGC\footnote{\url{https://github.com/RManLuo/NP-FKGC}} \cite{luo2023normalizing} integrates a stochastic ManifoldE encoder to incorporate the neural process and handle complex few-shot relations.
\end{itemize}
\textbf{Inductive KGC methods.} This group of methods can predict missing factual triples inductively.
\begin{itemize}
    \item PathCon\footnote{\url{https://github.com/hwwang55/PathCon}} \cite{wang2021relational} combines two types of subgraph structures (\textit{i.e.,} the contextual relations and the relational paths between entities) to predict link inductively.
    \item SNRI\footnote{\url{https://github.com/Tebmer/SNRI}} \cite{xu2022subgraph} leverages the complete neighbor relations of entities using extracted neighboring relational features and paths to handle sparse subgraphs.
    \item GraIL\footnote{\url{https://github.com/kkteru/grail}} \cite{teru2020inductive} is a GNN-based method to extract enclosing subgraph between two unseen entities for inductive link prediction.
    \item CSR\footnote{\url{https://github.com/snap-stanford/csr}} \cite{huang2022few} uses a connection subgraphs to represent entity pairs and test whether each triple in query set aligns with shared connection subgraph extracted from support set.
\end{itemize}

\subsection{Implementation Details}
We implement our model with PyTorch and PyG packages and conduct our experiments on a single RTX 3090 GPU. For the GNN encoder, we adopt a 3-layer PathCon with the hidden dimension set to 128. The dimension of subgraph representation and hypothesis $z$ is set to 100 for NELL as well as ConceptNet, and 50 for WIKI. The $\tau$ in stochastic attention is set to 0.7.
We use Adam \cite{kingma2014adam} as the optimizer. The negative sampling size is set to 1, the learning rate is set to $10^{-5}$, and the margin $\gamma$ is set to 1. The best model used for testing is selected by the MRR metric on the validation dataset. For baseline methods, we implement these methods from the repositories publicized by their authors. For the I-FKCG problem, we initialize entity embeddings randomly for our methods as they are unknown during testing. All methods use 100-dimensional relation and entities embedding for NELL and ConceptNet datasets and 50 dimensions for the WIKI dataset when applicable. 
Existing FKGC methods can generalize to new relations but not to unseen entities since the entity representations are constructed from a set of embeddings pretrained on background KGs. Metric learning methods such as GMatching, FSRL, and FAAN rely on encoding embeddings of entity pairs to measure the similarity between the support set and query set. Meta-optimization-based methods (\textit{e.g.,} MetaR, GANA) also require entity embeddings in the score function for prediction. However, under the I-FKGC setting where entities in the test phase are unseen, these methods experience significant performance degradation due to the lack of entity embeddings. Therefore, following previous studies \cite{huang2022few}, we violate the inductive setting and use pre-trained entity embeddings (\textit{e.g.,} TransE \cite{bordes2013translating}) for these FKGC methods to provide comparable results.

Inductive KGC methods are primarily designed for inductive settings and are not directly applicable to few-shot tasks. Consequently, adhering to their original configurations, we concatenate all meta-training tasks for the training phase and evaluate these methods across all query sets in the meta-testing phase for comparison.
%

\begin{figure*}[ht]
        \centering
        \begin{minipage}[t]{0.32\linewidth}
                \centering
                \includegraphics[width=\textwidth]{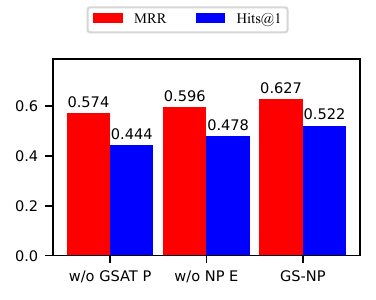}
                \label{fig:ablation1}
        \end{minipage}
        \begin{minipage}[t]{0.32\linewidth}
                \centering
                \includegraphics[width=\textwidth]{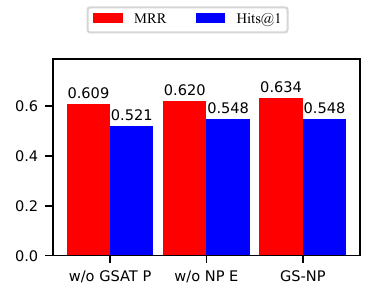}
                \label{fig:ablation2}
        \end{minipage}
        \begin{minipage}[t]{0.32\linewidth}
                \centering
                \includegraphics[width=\textwidth]{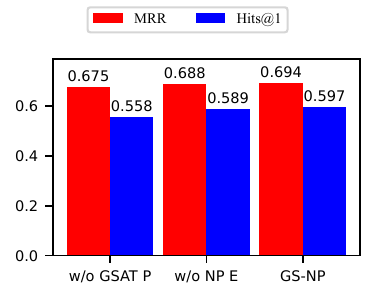}
                \label{fig:ablation3}
        \end{minipage}
        \captionsetup{justification=centering}
        \caption{Ablation study results on NELL, ConceptNet, and WIKI datasets.}
        \label{fig:ablation}
\end{figure*}

\begin{figure*}[ht]
        \centering
        \begin{minipage}[t]{0.32\linewidth}
                \centering
                \includegraphics[width=\textwidth]{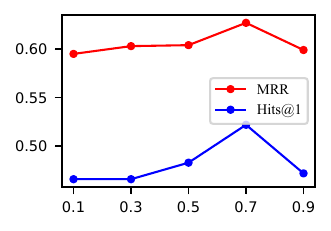}
                \label{fig:analysis4nell}
        \end{minipage}
        \begin{minipage}[t]{0.32\linewidth}
                \centering
                \includegraphics[width=\textwidth]{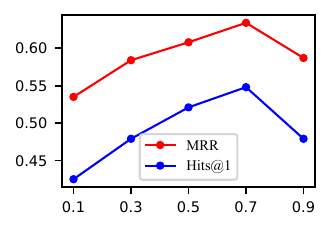}
                \label{fig:analysis4r2}
        \end{minipage}
        \begin{minipage}[t]{0.32\linewidth}
                \centering
                \includegraphics[width=\textwidth]{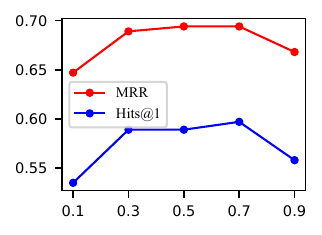}
                \label{fig:analysis4r3}
        \end{minipage}
        \captionsetup{justification=centering}
        \caption{Parameter analysis of $r$ on NELL, ConceptNet, and WIKI datasets.}
        \label{fig:analysis4r}
\end{figure*}

\subsection{Results and Analysis}
As shown in Table \ref{tab:main-table}, we report the results of 3-shot I-FKGC on the NELL, ConceptNet, and WIKI datasets. The best results are highlighted in bold and the second-best results are underlined. From the results, we can see that \ourmethod outperforms all baseline methods and derives new SOTA performance on most metrics, which demonstrates the effectiveness of our model.

Traditional inductive KGC methods get the worst results when compared with all baseline methods. Despite being designed for inductive reasoning, they overlook the few-shot settings, causing them to struggle with generalizing to unseen relations with few-shot triples. Early FKGC methods (\textit{e.g.,} GMatching, MetaR, FSRL) achieve better performance as they solve the problem in a meta-learning framework, which effectively utilizes the limited data to derive new knowledge. However, these methods often encounter out-of-distribution and overfitting issues, resulting in suboptimal results. To tackle this problem, NP-FKGC adopts the neural process to model the distribution of prediction functions and achieves improved performance. This demonstrates the effectiveness of the neural process in few-shot settings.

CSR casts the I-FKGC problem as an inductive reasoning problem, which attempts to extract the commonly shared graph structure as a hypothesis for predictions. As a result, CSR demonstrates good generalization to unseen entities and relations despite limited data, achieving the second-best performance among all baseline methods. Following the paradigm of inductive reasoning, \ourmethod also extracts the hypothesis from the support set by modeling the distribution of the hypothesis. By doing so, \ourmethod combines both the merits of neural process and inductive reasoning. Moreover, \ourmethod adopts a graph stochastic attention-based predictor to extract a label-relevant subgraph identified by the hypothesis for predictions. As a result, \ourmethod outperforms CSR, achieving the best performance among all baseline methods.

\begin{figure*}[t]
        \centering
        \begin{minipage}[t]{0.32\linewidth}
                \centering
                \includegraphics[width=\textwidth]{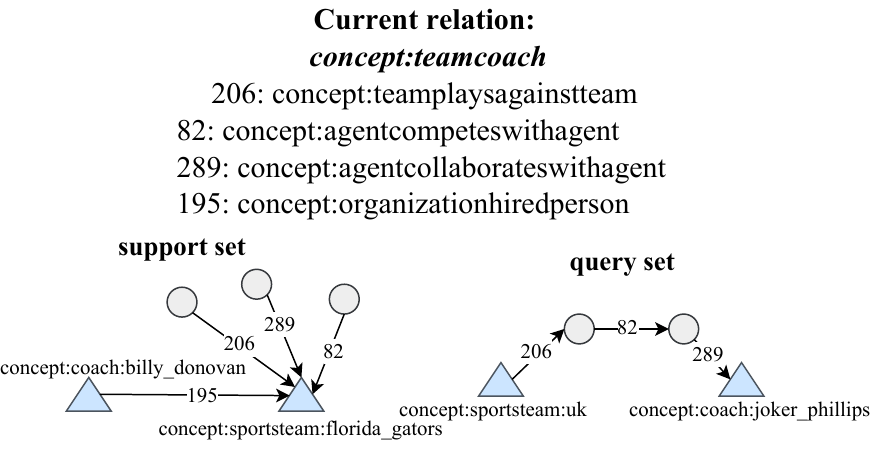}
                \caption{Illustration of extracted subgraph for a specific relation \textit{concept:teamcoach} in NELL dataset.}
                \label{fig:visualization}
        \end{minipage}
        \begin{minipage}[t]{0.32\linewidth}
                \centering
                \includegraphics[width=\textwidth]{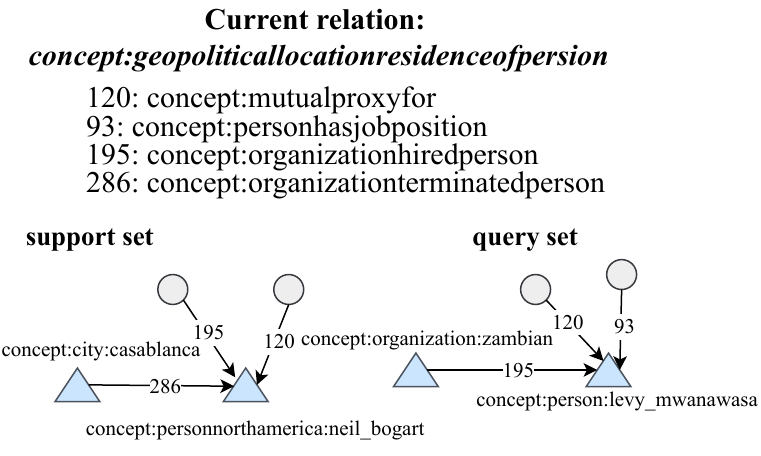}
                \caption{Illustration of extracted subgraph for \\a specific relation \textit{concept:geopoliticallocation\\residenceofpersion} in NELL dataset.}
                \label{fig:vis_1}
        \end{minipage}
        \begin{minipage}[t]{0.32\linewidth}
                \centering
                \includegraphics[width=\textwidth]{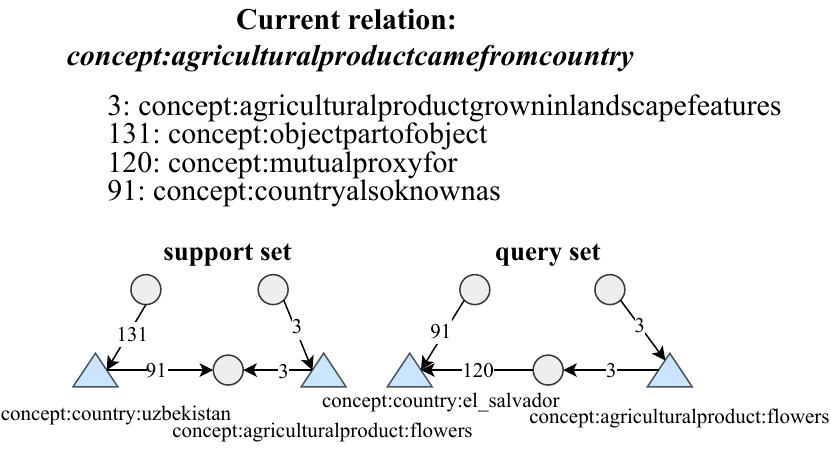}
                \caption{Illustration of extracted subgraph for \\a specific relation \textit{concept:agriculturalproduct\\camefromcountry} in NELL dataset.}
                \label{fig:vis_2}
        \end{minipage}
\end{figure*}

\subsection{Ablation Study}
To evaluate the effectiveness of the neural process-based hypothesis extractor (\textbf{NP Extractor}) and graph stochastic attention-based predictor (\textbf{GSAT Predictor}), we perform an ablation study by removing each component of \ourmethod. The experiments are conducted on the NELL, ConceptNet, and WIKI datasets with a 3-shot support set. From the results shown in Figure~\ref{fig:ablation}, where we abbreviate \textbf{NP Extractor} and \textbf{GSAT Predictor} as \textbf{NP E} and \textbf{GSAT P}, we can observe that all components can help improve the performance of our model. By removing the NP extractor, the extracted hypothesis is the average graph representation of the support set, which may not generalize well to the query set. Without the GSAT Predictor, the performance drops a lot as the extracted hypothesis is a latent variable. Applying it directly to graph structure impairs the performance and lacks explanations.



\subsection{Parameter Analysis}
\label{sec:pa}
We study the impact of the selection of $\tau$ in Eq.~\eqref{eq:r}. From Figure~\ref{fig:analysis4r}, we can see that results decrease as $\tau$ decreases to 0. The best performance is achieved when $\tau$ is around 0.7. This is consistent with the theory observed by GSAT \cite{miao2022interpretable} that sets $\tau$ to a value close to 1 can extract a much denser subgraph and often provide a more robust interpretation.

\begin{figure}[t]
        \centering
        \begin{minipage}[t]{0.49\columnwidth}
                \centering
                \includegraphics[width=\textwidth]{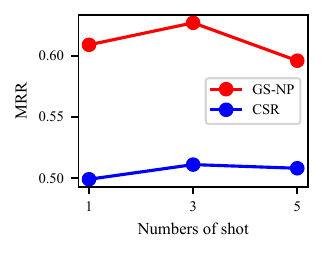}
                \caption{MRR metric under different $K$-shot settings on NELL dataset.}
                \label{fig:fewshot1}
        \end{minipage}
        \begin{minipage}[t]{0.49\columnwidth}
                \centering
                \includegraphics[width=\textwidth]{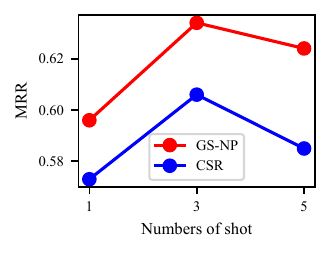}
                \caption{MRR metric under different $K$-shot settings on ConceptNet dataset.}
                \label{fig:fewshot2}
        \end{minipage}
\end{figure}

To further study the ability of inductive reasoning, we conduct experiments under different $K$-shot scenarios compared to the existing SOTA method CSR. Figure~\ref{fig:fewshot1} and Figure~\ref{fig:fewshot2} illustrate the MRR metric under different $K$-shot scenarios on the NELL and ConceptNet datasets. As the value of $K$ decreases to 1, the performance of CSR and \ourmethod decline, and \ourmethod is better than CSR. This suggests that by modeling the distribution of the hypothesis, our method can perform well even with just one triple. However, as $K$ increases to 5, the performance of CSR and our method drops. One possible explanation for this is that when more triples in the support set are given, noise is introduced, which impairs the performance of both CSR and our method.

\subsection{A Study on Transductive Setting}
To further demonstrate the effectiveness of our method, we also test the performance of \ourmethod under a transductive setting \cite{9459451} in the NELL and ConceptNet datasets, where the entities are seen during training. For baseline methods, we directly use the results reported by CSR \cite{huang2022few}. From the results shown in Table~\ref{tab:trans}, we can observe that \ourmethod can also be applied to transductive settings and achieve the best performance among all baseline methods. This is because \ourmethod can extract meaningful hypotheses from subgraphs which can effectively be used to derive new knowledge.

\begin{table}[t]
        \centering
        \caption{Results of 3-shot transductive FKGC on NELL and ConceptNet datasets.}
        \label{tab:trans}
        \resizebox{\columnwidth}{!}{%
                \begin{tabular}{p{2cm}|cccc|cccc}   
                        \toprule
                             \multirow{2}{*}{Method}      & \multicolumn{4}{c|}{\textbf{NELL}} & \multicolumn{4}{c}{\textbf{ConceptNet}} \\ \cmidrule(r){2-5} \cmidrule(r){6-9} 
                                   & \multicolumn{1}{c}{\textbf{MRR}}  & \multicolumn{1}{c}{\textbf{Hit@1}}      & \multicolumn{1}{c}{\textbf{Hit@5}} & \multicolumn{1}{c|}{\textbf{Hit@10}} & \multicolumn{1}{c}{\textbf{MRR}} & \multicolumn{1}{c}{\textbf{Hit@1}} & \multicolumn{1}{c}{\textbf{Hit@5}} & \multicolumn{1}{c}{\textbf{Hit@10}} \\ \midrule 

                        MetaR      & 0.471          & 0.322            & 0.647            & 0.763        &0.318      &0.226      &0.390      &0.496    \\ 
                        FSRL       & 0.490          & 0.327            & 0.695            & 0.853        &0.577      &0.469      &0.695      &0.753    \\ 
                        NP-FKGC    & 0.509          & 0.388            & 0.635            & 0.736        &0.265      &0.139      &0.316      &0.644    \\ 
                        CSR        & 0.577          & 0.442            & 0.746            & \textbf{0.858} &0.606      &0.495      &\textbf{0.735}  &\textbf{0.777} \\ \midrule 
                        \ourmethod & \textbf{0.617} & \textbf{0.495}   & \textbf{0.771}   & 0.857       &\textbf{0.608}      &\textbf{0.512}      &0.720      &0.755   
                        \\ \bottomrule
                \end{tabular}%
        }
\end{table}


\subsection{Explanatory Subgraph for Predictions}
The stochastic attention mechanism in \ourmethod can generate an edge mask to extract a subgraph identified by the hypothesis. The extracted subgraph visualizes the hypothesis and explains the predictions. In Figure~\ref{fig:visualization}, we illustrate the extracted subgraph for a specific relation \texttt{concept:teamcoach} in the NELL dataset. From the visualization, it is evident that our method extracts a meaningful hypothesis from the support set, \textit{e.g.,}
\texttt{concept:teamplaysagainstteam}, \texttt{concept:ag\\entcompeteswithagent}, \texttt{concept:agentcollab\\orateswithagent} $\Rightarrow$ \texttt{concept:teamcoach}. All three relations in the hypothesis can be found in the subgraph of the support set, indicating that our method effectively captures structure dependence from limited data.
Based on the hypothesis, we can infer a potential relation \texttt{concept:teamcoach} between the head entity \texttt{concept:sportsteam:uk} and the tail entity \texttt{concept:sportsteam:uk} in the query set. 

As shown in Figure~\ref{fig:vis_1}, we illustrate the relation \texttt{concept:geopoliticallocationresidenceofper\\sion} while as shown in Figure~\ref{fig:vis_2}, we illustrate the relation \texttt{concept:agriculturalproductcamefromcountry}.



\section{Conclusion}
\label{sec:con}
In this paper, we propose a novel graph stochastic neural process approach, called \ourmethod for inductive few-shot knowledge graph completion (I-FKGC). Our method is motivated by the concept of inductive reasoning. To extract a general hypothesis, we employ a neural process-based hypothesis extractor to capture the structure dependence and model the joint distribution of the hypothesis, based on which we obtain a hypothesis to represent the shared pattern within the support set. We then use a graph stochastic attention-based predictor to inject the extracted hypothesis into the graph structure and identify a subgraph from the query for the predictions. In this way, we can not only test the applicability of the hypothesis but also provide explanatory predictions. Extensive experiments on three benchmark datasets demonstrate the effectiveness of our method.

In the future, we plan to unify large language models (LLMs) and knowledge graphs to improve the performance for inductive few-shot knowledge graph completion \cite{pan2024unifying}, \cite{9547420}. Specifically, we will utilize the LLMs to enrich representations of KGs by encoding the textual descriptions of entities and relations, which facilitates the downstream tasks \cite{chen2024exploring}. Meanwhile, we will also explore the potential of KGs to enhance LLMs' faithful and interpretable reasoning on complex tasks \cite{luo2023reasoning,pan2024integrating}. Besides, we will try to explore the potential of \ourmethod for temporal knowledge graph completion tasks.
\section*{Appendix}
\setcounter{section}{0}
\renewcommand{\thesection}{\Alph{section}}
\section{Derivation of ELBO Loss}
\label{sec:elbo}
\setcounter{equation}{0}
\renewcommand{\theequation}{A.\arabic{equation}}
Given a relation $r_q$ and its associated support set $\mathcal{C}_{r_q}$, the objective of \ourmethod is to infer the distributions $P(Z | \mathcal{C}_{r_q}, \mathcal{G}_q)$ from the support set that minimizes the prediction loss on the query set $\mathcal{D}_{r_q}$. The unified training objective function for \ourmethod can be formulated as
\begin{equation}
    \begin{aligned}
        &P\left(t_q, Z | h_q, r_q, \mathcal{C}_{r_q}, \mathcal{G}_q\right) \\ 
        &= P(Z | \mathcal{C}_{r_q}, \mathcal{G}_q) \prod_{\{(h_q,r_q,?)\}} P(t_q|f_{r_q}(h_q,r_q,Z)),
        \label{eq_1}
    \end{aligned}
\end{equation}
where $f_{r_q}(h_q,r_q,Z)$ denotes the \textit{predictor}, $\{(h_q,r_q,?)\}$ denotes the query to be predicted, and $Z=(\mathcal{G}_q, z)$ denotes all the latent variables. 

Following the Eq.~\eqref{eq_1}, the prediction likelihood on the query set $P(t_q | h_q, r_q, \mathcal{C}_{r_q}, \mathcal{G}_q)$ can be written as
\begin{equation}
\begin{aligned}
    \log P(t_q | h_q, r_q, \mathcal{C}_{r_q}, \mathcal{G}_q) = \log \frac{P\left(t_q, Z | h_q, r_q, \mathcal{C}_{r_q},\mathcal{G}_q\right)}{P\left(Z | \mathcal{C}_{r_q},\mathcal{G}_q\right)} \\
    = \log P\left(t_q, Z | h_q, r_q, \mathcal{C}_{r_q},\mathcal{G}_q\right) - \log P\left(Z | \mathcal{C}_{r_q},\mathcal{G}_q\right).
    \label{eq_2}
\end{aligned}
\end{equation}

Assuming that $Q(Z)$ is the true distribution of $Z$, we can rewrite the Eq.~\eqref{eq_2} as
\begin{equation}
\begin{aligned}
    &\log P(t_q | h_q, r_q, \mathcal{C}_{r_q}, \mathcal{G}_q) \\
    &= \log \frac{P\left(t_q, Z | h_q, r_q, \mathcal{C}_{r_q},\mathcal{G}_q\right)}{Q(Z)} - \log \frac{P\left(Z | \mathcal{C}_{r_q},\mathcal{G}_q\right)}{Q(Z)}.
    \label{eq_3}
\end{aligned}
\end{equation}

Then, we integrate both sides with $Q(Z)$, and obtain
\begin{equation}
\resizebox{\columnwidth}{!}{$
\begin{aligned}
    \label{eq_4}
    &\log P(t_q | h_q, r_q, \mathcal{C}_{r_q}, \mathcal{G}_q) \\
    = &\int_Z Q(Z) \log \frac{P\left(t_q, Z | h_q, r_q, \mathcal{C}_{r_q},\mathcal{G}_q\right)}{Q(Z)} - \int_Z Q(Z) \log \frac{P\left(Z | \mathcal{C}_{r_q},\mathcal{G}_q\right)}{Q(Z)} \\
    = & \int_Z Q(Z) \log \frac{P\left(t_q, Z | h_q, r_q, \mathcal{C}_{r_q},\mathcal{G}_q\right)}{Q(Z)} + KL \left(Q(Z) \| P\left(Z | \mathcal{C}_{r_q},\mathcal{G}_q\right)\right).
\end{aligned}$
}
\end{equation}

Since $ KL(Q(Z) \| P\left(Z | \mathcal{C}_{r_q},\mathcal{G}_q\right) \ge 0$, we can derive Eq.~\eqref{eq_4} as
\begin{equation}
\resizebox{\columnwidth}{!}{$
\begin{aligned}
    &\log P(t_q | h_q, r_q, \mathcal{C}_{r_q}, \mathcal{G}_q) \ge \int_Z Q(Z) \log \frac{P\left(t_q, Z | h_q, r_q, \mathcal{C}_{r_q},\mathcal{G}_q\right)}{Q(Z)} \\
    &= \mathbb{E}_{Q(Z)} \log \frac{P\left(t_q,Z|h_q, r_q, \mathcal{C}_{r_q}, \mathcal{G}_q\right)}{Q(Z)} \\
    &= \mathbb{E}_{Q(Z)}\left[\log P\left(t_q | h_q, r_q, Z\right)+\log \frac{P\left(\mathcal{G}_S,z | h_q, r_q, \mathcal{C}_{r_q}, \mathcal{G}_q\right)}{Q(Z)}\right] \\
    & =\mathbb{E}_{Q(Z)}\left[\log P\left(t_q | h_q, r_q, Z\right) + \log \frac{P\left(z | \mathcal{C}_{r_q}\right)}{Q(z)} + \log \frac{P\left(\mathcal{G}_S | h_q, r_q, z, \mathcal{G}_q\right)}{Q(\mathcal{G}_S)}\right] \\
    & =\underbrace{\mathbb{E}_{Q(Z)}\left[\log P\left(t_q | h_q, r_q, Z\right)\right]}_{\text{(1)}}-\underbrace{K L\left(Q(z) \| P\left(z | \mathcal{C}_{r_q}\right)\right)}_{\text{(2)}}                                                          \\
    & -\underbrace{KL\left(Q\left(\mathcal{G}_S\right) \| P\left(\mathcal{G}_S | \mathcal{G}_q, z\right)\right)}_{\text{(3)}}.
    \label{eq_5}
\end{aligned}
$}
\end{equation}

Therefore, we obtain the unified objective function as presented in Eq.~\eqref{eq:objective}, enabling simultaneous optimization of the Neural Process-based Hypothesis Extractor and the Graph Stochastic Attention-based Predictor.

\bibliographystyle{IEEEtran}
\bibliography{main.bib}

\end{document}